\documentclass[3p,twocolumn,sort&compress, fleqn]{elsarticle}
\usepackage{amsmath, bm}
\usepackage{amssymb}
\usepackage{amsmath}
\usepackage{braket}
\usepackage{listings}

\journal{Computer Physics Communications}

\def\mathi{\mathrm i}

\def\deltaB{\delta_{\alpha, \mathrm{B}}}
\def\deltaF{\delta_{\alpha, \mathrm{F}}}
\def\G1{{G^{(1)}}}

\newcommand{\KF}{\ensuremath{K^\mathrm{F}}}
\newcommand{\KB}{\ensuremath{K^\mathrm{B}}}
\newcommand{\kF}{\ensuremath{k^\mathrm{F}}}
\newcommand{\kB}{\ensuremath{k^\mathrm{B}}}

\newcommand{\wmax}{\ensuremath{{\omega_\mathrm{max}}}}

\usepackage{natbib}
\usepackage{hyperref}
\hypersetup{
        colorlinks=true,
}
\usepackage{listings}

\long\def\beginmypgfpdfnamed#1#2\endmypgfpdfnamed{\includegraphics{#1}}

\graphicspath{{./pyfig/}}

\usepackage[whole]{bxcjkjatype}

\begin{document}

\begin{frontmatter}

\title{irbasis: Open-source database and software for intermediate-representation basis functions of imaginary-time Green's function}

\author[saitama]{Naoya Chikano}
\author[issp]{Kazuyoshi Yoshimi}
\author[tohokuuniv]{Junya Otsuki\fnref{okayama}}
\author[saitama]{Hiroshi Shinaoka} \ead{shinaoka@mail.saitama-u.ac.jp}

\address[saitama]{Department of Physics, Saitama University, Saitama 338-8570, Japan}
\address[issp]{Institute for Solid State Physics, University of Tokyo, Chiba 277-8581, Japan}
\address[tohokuuniv]{Department of Physics, Tohoku University, Sendai 980-8578, Japan}
\fntext[okayama]{Present address: Research Institute for Interdisciplinary Science, Okayama University, Okayama 700-8530, Japan}

\begin{abstract}
The open-source library, irbasis, provides easy-to-use tools for two sets of orthogonal functions named intermediate representation (IR).
The IR basis enables a compact representation of the Matsubara Green's function and efficient calculations of quantum models.
The IR basis functions are defined as the solution of an integral equation whose analytical solution is not available for this moment.
The library consists of a database of pre-computed high-precision numerical solutions and computational code for evaluating the functions from the database.
This paper describes technical details and demonstrates how to use the library.
\end{abstract}

\begin{keyword}
Matsubara/imaginary-time Green's function, many-body quantum theories
\end{keyword}

\end{frontmatter}
{\bf PROGRAM SUMMARY}

\begin{small}
\noindent
{\em Program Title:} irbasis \\
{\em Journal Reference:}                                      \\
{\em Catalogue identifier:}                                   \\
{\em Licensing provisions:} MIT license\\
{\em Programming language:} \verb*#C++#, Python. \\
{\em Computer:} PC, HPC cluster \\ 
{\em Operating system:} Any, tested on Linux and Mac OS X\\ 
{\em RAM:} less than 100 MB.\\ 
{\em Number of processors used:} 1.\\ 
{\em Keywords:} Imaginary-time Green's function, Matsubara Green's function\\ 
{\em Classification:} 4.4  \\ 
{\em External routines/libraries:}  numpy, scipy and h5py for Python library, HDF5 C library for \verb*#C++# library. \\
{\em Nature of problem:} Numerical orthogonal systems for Green's function\\
{\em Solution method:} Galerkin method, piece-wise polynomial representation\\
{\em Running time:} $<$ 1 min\\
\end{small}

\section{Introduction}
In condensed matter physics, 
Matsubara Green's function techniques are  powerful tools to study many-body physics in strongly correlated systems.
Examples include diagrammatic expansions such as the random phase approximation (RPA), the dynamical mean-field theory (DMFT)~\cite{Georges:1996un,0295-5075-100-6-67001}, GW method~\cite{Hedin:1965tu,doi:10.1021/ct300648t,0034-4885-61-3-002}), continuous-time quantum Monte Carlo (QMC) methods~\cite{Rubtsov:2005iwa,Werner:2006ko,Gull:2008cma,Otsuki:2007ff,Gull:2011jda}.

In practical calculations, the data of the Green's function is stored in computer memory.
When the Green's function is represented in the Matsubara-frequency domain as $G(i\omega_n)$, it exhibits a power-law decay at high frequencies.
Thus, the number of Matsubara frequencies required for representing $G(i\omega_n)$ grows rapidly with decreasing temperature.
Especially when the Green's function has other indices as spin, orbit, and wavenumber, the data of $G(i\omega_n)$ becomes huge at low temperatures.
Therefore, there is a high demand for a compact representation of the imaginary-time dependence of the Green's function in practical calculations.
Orthogonal polynomial representations of the Green's function are a common way to represent the data compactly~\cite{Boehnke2011,Gull2018}.

\begin{figure}
	\centering
	\includegraphics[width=0.5\textwidth,clip]{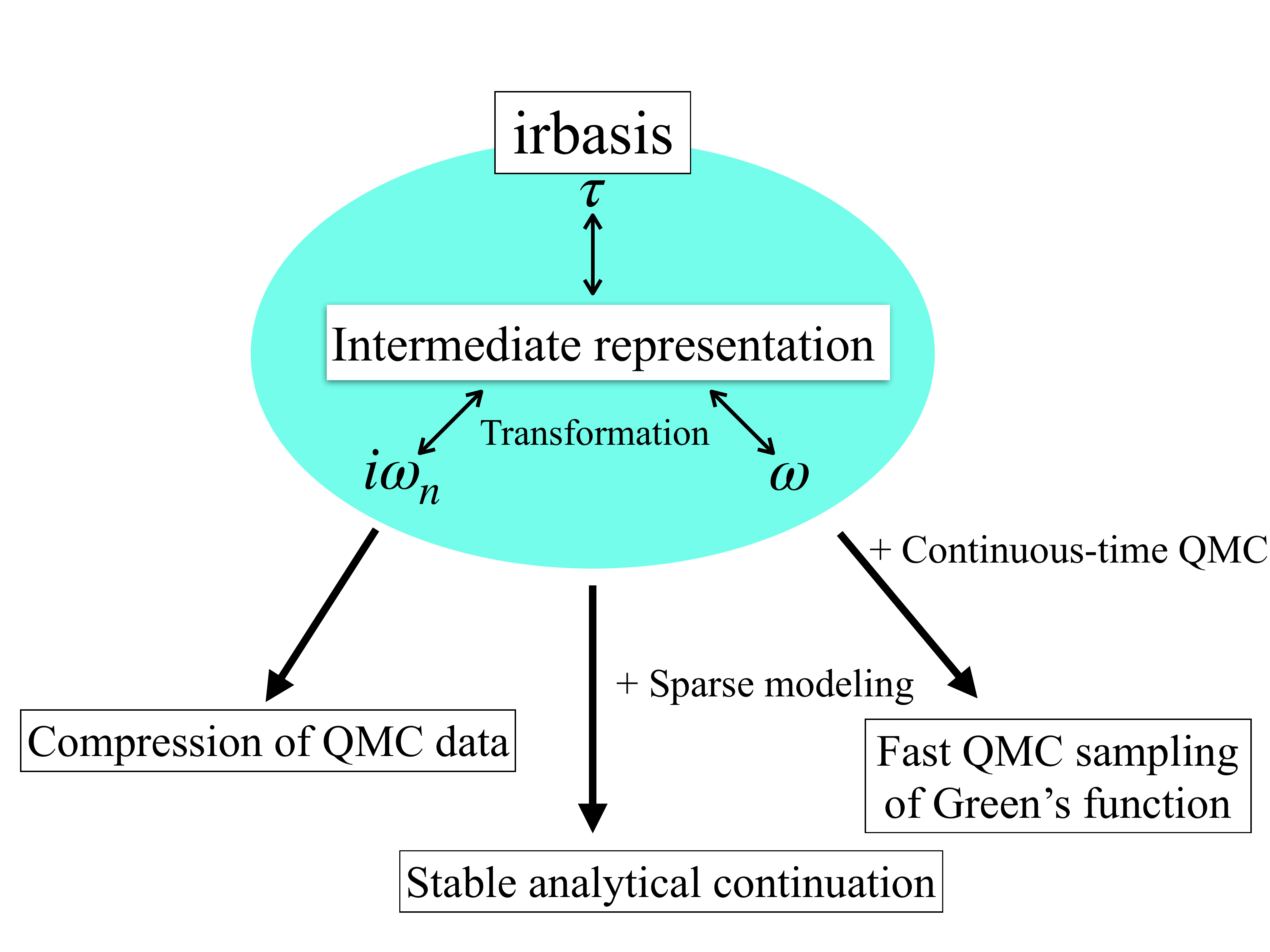}
	\caption{
		(Color online) Overview of the functionality of irbasis and applications.
	}
	\label{fig:schematic}
\end{figure}

Recently, it was found that there is a physical complete basis set which yields a much more compact representation of the Green's function~\cite{Shinaoka2017,Shinaoka2018,Chikano2018} than the conventionally used classical orthogonal polynomials such as Legendre/Chebyshev polynomials.
This physical representation was named the \textit{intermediate representation} (IR).
As illustrated in Fig.~\ref{fig:schematic}, the IR has been applied to a stable analytical continuation of QMC data to real frequencies in conjunction with sparse modeling techniques in data science~\cite{Otsuki:2017er},
fast QMC sampling of the single-/two-particle Green's function~\cite{Shinaoka2017},
and the analysis of the self-energy computed by DMFT calculations with exact-diagonalization techniques~\cite{Nagai2018}.
The use of the IR basis functions thus will open up new and interesting research applications in condensed matter physics.

\begin{figure}
	\centering
	\includegraphics[width=0.4\textwidth,clip]{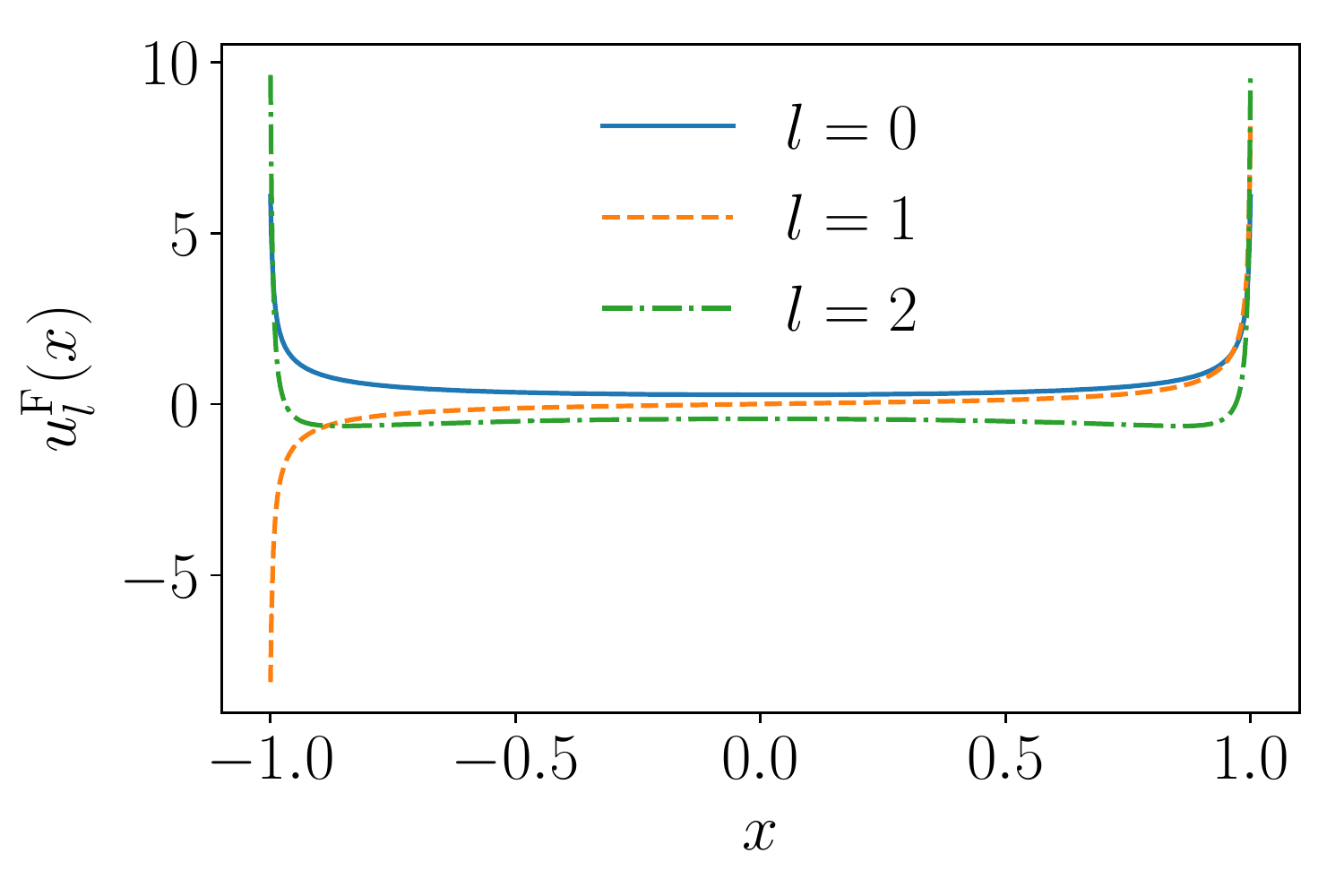}
	\includegraphics[width=0.4\textwidth,clip]{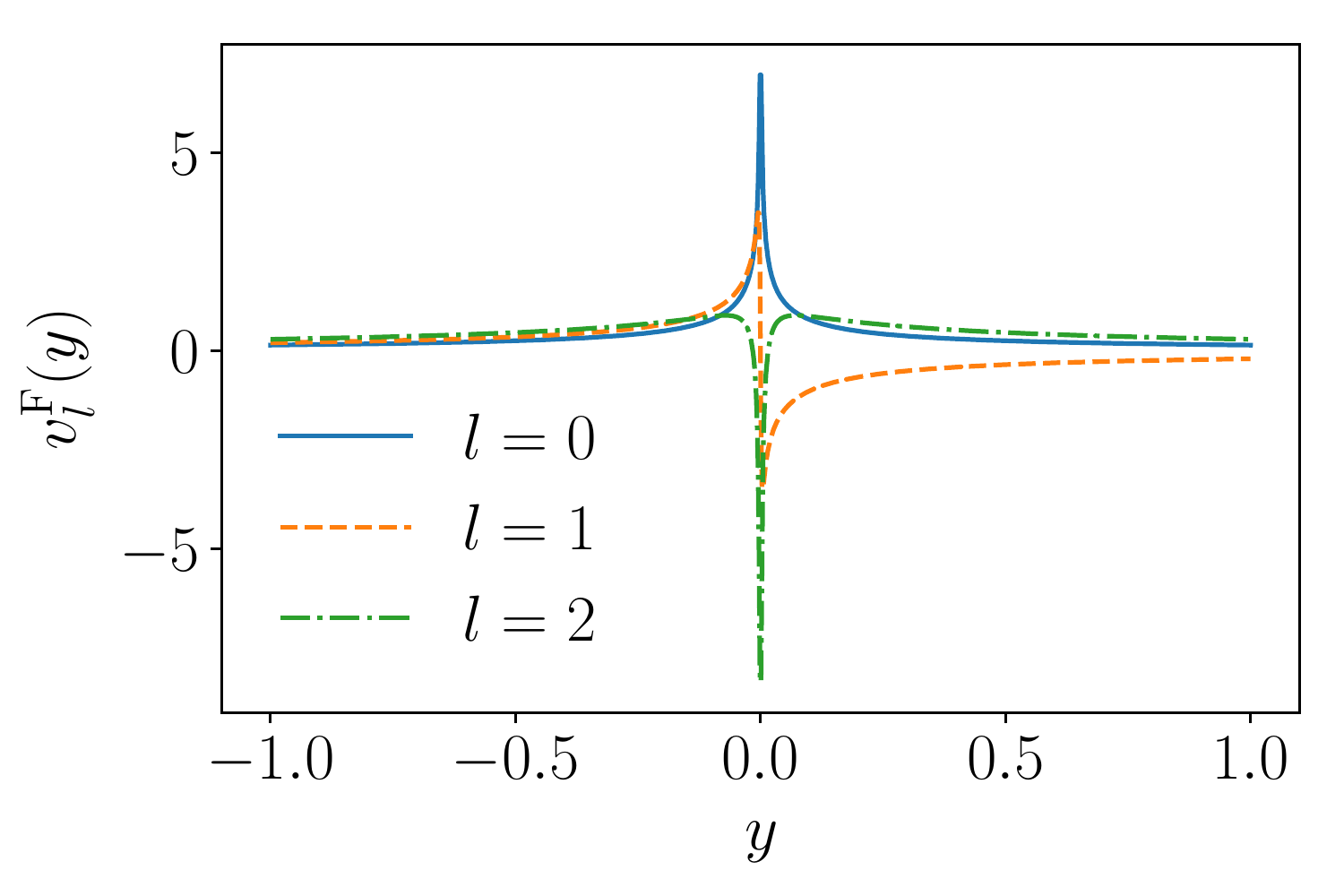}
	\caption{
		(Color online) IR basis functions $u_l^\alpha(x)$ and $v_l^\alpha(y)$ for $\Lambda = 1000$ and fermions.
		$x$ and $y$ are dimensionless variables corresponding to imaginary time and real frequency, and $l$ enumerates the basis functions.
		These two sets of basis functions are orthonormalized with a constant weight function of 1 in $[-1,1]$, respectively.
	}
	\label{fig:uv}
\end{figure}

Figure~\ref{fig:uv} shows a typical example of the IR basis functions.
The basis consists of a pair of two \textit{complete} and \textit{orthogonal} basis sets $\{u_l^\alpha(x)\}$, $\{v_l^\alpha(y)\}$ ($x$, $y$ correspond to imaginary time and real frequency as described later).
They depend on the statistics $\alpha$ (fermions or bosons) and a dimensionless parameter $\Lambda > 0$ (the definition is given later).
They involve Legendre polynomials as a special limit of $\Lambda = 0$.
The IR basis functions share favorable mathematical properties (e.g., orthogonality) with classical orthogonal polynomials.
Thus, the IR basis can be used in many applications. However, \textit{one practical problem remains to be solved as described below.}

There exist many numerical libraries for evaluating classical orthogonal polynomials.
The evaluation is immediate thanks to their recurrence relations.
On the other hand, the IR basis functions are the solution of an integral equation whose analytical solution is not available for the moment.
For applications ranging from QMC to diagrammatic calculations,
the basis functions must be determined typically within a relative error of $10^{-8}$.
Recently, some of the authors developed a method for computing a precise numerical solution of the integral equation~\cite{Chikano2018}.
To reach such precision, however, the algorithm must be implemented in arbitrary-precision arithmetic because the integral equation is ill-conditioned.
This makes the computation very expensive and requires the use of a arbitrary-precision library such as GMP.
Furthermore, a special care must be taken to avoid discretization errors in representing the rapidly oscillating basis functions (see Fig.~\ref{fig:uv}).

Thus, there is still missing an easy-to-use and computationally cheap numerical library for evaluating the IR basis functions.
Our \textit{irbasis} library eliminates this bottleneck
by providing a database of pre-computed precise numerical solutions and a simple and fast tool for interpolating basis functions.
The library is implemented in \verb*#C++# and Python with minimal dependencies on external libraries to ensure its portability.
The interpolation is performed in double-precision arithmetic but with a controlled accuracy, being computationally inexpensive.
These features will enable to use IR basis functions as easily as classical orthogonal polynomials in many practical applications without technical difficulties.

The remainder of this paper is organized as follows.
In Section 2, we review the definition of IR basis functions and establish notations.
We describe the structure of irbasis library in Section 3.
The installation and basic usage are described in Sections 4 and 5, respectively.
Step-by-step examples in Section 6 provide an explanation of how to use the library for many-body calculations.
Section 7 presents a summary.

\section{Intermediate representation (IR) basis function}
\subsection{Definition}
We briefly introduce the definition of the IR basis functions following the previous studies with a slight modification of notations.
We start with the spectral (Lehmann) representation of the single-particle Green’s function $G^{\alpha}(\tau)$ in the imaginary-time domain
\begin{equation}
G^{\alpha}(\tau) = -\int^{\wmax}_{-\wmax}d\omega K^{\alpha}(\tau,\omega)\rho(\omega)\label{eq:spectral}
\end{equation}
where $\beta$ is the inverse temperature and we assume $\hbar = 1$.
The superscript $\alpha$ specifies statistics: $\alpha=\mathrm{F}$ for fermion and $\alpha=\mathrm{B}$ for boson.
The spectral function is defined as
 \begin{align}
 \rho^\alpha(\omega) &= -\frac{1}{\pi\omega^{\delta_{\alpha,\mathrm{B}} }} \mathrm{Im} G^\alpha(\omega + \mathi 0).\label{eq:rho}
 \end{align}
We assume that the spectrum $\rho^{\alpha}(\omega)$ is bounded in the interval $[-\wmax, \wmax]$,
where $\wmax$ is a cutoff frequency.
The kernel $K^{\alpha}(\tau,\omega)$ reads
\begin{align}
K^\alpha(\tau, \omega) &\equiv \omega^{\delta_{\alpha, \mathrm{B}}} \frac{e^{-\tau\omega}}{1 \pm e^{-\beta \omega}}.\label{eq:K}
\end{align}
for $0 \leq \tau \leq \beta$.
Here, the $+(-)$ sign is for fermions and bosons, respectively.
The extra $\omega$'s for boson in Eqs.~(\ref{eq:rho}) and (\ref{eq:K}) was introduced to avoid a singularity of the kernel at $\omega=0$.

For given $\wmax$ and $\beta$,
the IR basis functions are defined through the decomposition
\begin{equation}
K^\alpha(\tau,\omega) = \sum^{\infty}_{l=0} S^\alpha_l U^{\alpha}_l(\tau) V^{\alpha}_l(\omega),\label{eq:decomp-tau-omega}
\end{equation}
where $\int_0^\beta d \tau U_l^\alpha(\tau)U_{l^\prime}^\alpha(\tau) = \int_{-\wmax}^\wmax d \omega V_l^\alpha(\omega)V_{l^\prime}^\alpha(\omega) = \delta_{ll^\prime}$.
Taking $U_l^\alpha(\beta) >0$, we fix the sign of $U^\alpha_l(\tau)$ and $V^\alpha_l(\omega)$.
This decomposition corresponds to the singular value decomposition (SVD) of the matrix representation of the kernel defined on a discrete $\tau$-$\omega$ space.
Singular values $S_l^\alpha$ ($>0$) are given in decreasing order.
Note that
$U_l^\alpha(\tau) = (-1)^l U_l^\alpha(\beta-\tau)$ and $V_l^\alpha(\omega) = (-1)^l V_l^\alpha(-\omega)$ for $0 < \tau < \beta$.
The basis functions and singular values can be computed by solving the integral equation
\begin{align}
	S^\alpha_l U^\alpha_l(\tau) &= \int_{-\wmax}^\wmax d\omega K^\alpha(\tau, \omega) V^\alpha_l(\omega).\label{eq:integral-UV}
\end{align}

The imaginary-time Green's function and the corresponding spectral function can be expanded as
\begin{align}
G^{\alpha}(\tau) &= \sum_{l=0}^\infty  G_l^{\alpha} U_l^{\alpha}(\tau)\label{eq:IR-decomp},\\
\rho^\alpha(\omega) &= \sum_{l=0}^\infty \rho_l^\alpha V^\alpha_l(\omega),\label{eq:rhol}
\end{align}
where
\begin{align}
G_l^\alpha &= - S_l^\alpha \rho_l^\alpha.\label{eq:Gl_rhol}
\end{align}

The orthogonality of basis functions yields the inverse transform of Eqs.~(\ref{eq:IR-decomp}) and (\ref{eq:rhol}) as
\begin{align}
G_l^\alpha &= \int^{\beta}_{0}d\tau G^{\alpha}(\tau)U_l^{\alpha}(\tau) \label{eq:Gl}\\
& =
-S^{\alpha}_l \int^{\wmax}_{-\wmax}d\omega \rho^{\alpha}(\omega)V_l^{\alpha}(\omega),\\
\rho^\alpha_l &= \int^{\wmax}_{-\wmax}d\omega \rho^\alpha(\omega) V^\alpha_l(\omega).\label{eq:Gl_rhol2}
\end{align} 
Note that a constant term in $G^\mathrm{B}(\tau)$ is not represented by the IR basis compactly, and thus should be treated separately (refer to \ref{appendix:pole} and Ref.~\cite{Chikano2018}).

A striking feature of this decomposition is the exponential decay of $S_l^{\alpha}$.
Although there is no available analytic proof,
extensive numerical investigation in previous studies supports the exponential decay regardless of $\wmax$ and $\beta$~\cite{Shinaoka2017, Chikano2018}.
This guarantees that $G_l^\alpha$ decay fast no matter how fast $\rho^\alpha_l$ decay.

The Matsubara-frequency representation of the Green's function can be obtained by the Fourier transformations
\begin{align}
G^\alpha(i\omega_n) &= \mathcal{F}(G^\alpha(\tau)) = \int_0^\beta d \tau G^\alpha(\tau) e^{i\omega_n \tau} \nonumber\\
& = \sum_{l=0}^\infty G_l^\alpha U^\alpha_l(i\omega_n)\label{eq:Gl-omega}\\
&\mathrm{with}~U^\alpha_l(i\omega_n) \equiv \int_0^\beta d \tau U^\alpha_l(\tau) e^{i\omega_n \tau},
\end{align}
where $U_l(i\omega_n)$ is the Fourier transformation of $U(\tau)$.
$\mathcal{F}$ is the Fourier transformation operator.

Equation~(\ref{eq:spectral}) can be reformulated as
\begin{align}
	G^{\alpha}(i\omega_n) &= \int_{-\infty}^{\infty} d\omega K^{\alpha}(i\omega_n, \omega) \rho^{\alpha}(\omega),\label{eq:fwd-iwn}
\end{align}
where
\begin{align}
	\KF(i\omega_n, \omega) &\equiv -\mathcal{F}(\KF(\tau, \omega)) = \frac{1}{i\omega_n - \omega},\\
	\KB(i\omega_n, \omega) &\equiv -\mathcal{F}(\KB(\tau, \omega)) =\frac{\omega}{i\omega_n - \omega}.\label{eq:KB}
\end{align}
The decomposition of $K^\alpha(i\omega_n, \omega)$ reads
\begin{align}
    K^\alpha(i\omega_n, \omega) &= - \sum_{l=0}^\infty S_l U_l(i\omega_n) V_l(\omega).\label{eq:Komega-decomp}
\end{align}

\subsection{Dimensionless representation}
\label{sec:dimensionless}
In practical implementation,
it is convenient to use the dimensionless form of IR basis function $u_l^\alpha(x)$ and $v_l^\alpha(y)$ proposed in Ref.~\cite{Shinaoka2017}.
The dimensionless form of the basis functions $u^\alpha_l(x)$ and $v^\alpha_l(y)$ and singular values $s_l^\alpha$ are defined by the decomposition
\begin{align}
k^\alpha(x, y) = \sum^{\infty}_{l=0} s^\alpha_l u^{\alpha}_l(x) v^{\alpha}_l(y)\label{eq:decomp-x-y}
\end{align}
in the intervals of $x \in [-1,1]$ and $y \in [-1,1]$.
The basis functions $u_l^\alpha(x)$ and $v^\alpha_l(y)$ are orthonormalized in these intervals, respectively.
Here, the dimensionless form of the kernels is defined as
\begin{align}
\kF(x, y)&\equiv \frac{e^{-\frac{\Lambda}{2} x y}}{2\cosh(\frac{\Lambda}{2} y)},\label{eq:kernel-F-xy}\\
\kB(x, y)&\equiv y \frac{e^{-\frac{\Lambda}{2} x y}}{2\sinh(\frac{\Lambda}{2} y)},\label{eq:kernel-B-xy}
\end{align}
where $\Lambda$ is a dimensionless parameter.
We take $u_l^\alpha(1)>0$, which also fixes the sign of $v_l^\alpha(y)$.
Note that $u_l^\alpha(x) = (-1)^l u_l^\alpha(x)$ and $v_l^\alpha(y) = (-1)^l v_l^\alpha(-y)$.
They satisfy
\begin{align}
	s^\alpha_l u^\alpha_l(x) &= \int_{-1}^1 dy k^\alpha(x, y)v^\alpha_l(y).\label{eq:integral-xy}
\end{align}
We should note that the dimensionless form of the basis functions depend on only one parameter, $\Lambda$.

Comparing Eqs.~(\ref{eq:integral-UV}) and (\ref{eq:decomp-x-y}) establishes the relations between the two representations
\begin{align}
\Lambda &= \beta \wmax,\label{eq:Lambda}\\
U_l^\alpha(\tau) &= \sqrt{\frac{2}{\beta}} u_l^\alpha(x(\tau)),\label{eq:Ul}\\
V_l^\alpha(\omega) &= \sqrt{\frac{1}{\wmax}} v_l^\alpha(y(\omega)),\label{eq:Vl}\\
 S^\alpha_l &= \sqrt{\frac{\beta\wmax^{1+2\deltaB}}{2}}s^\alpha_l, \label{eq:sl}\\
U^\alpha_l(i\omega_n) &= \sqrt{\beta} u_{nl}^\alpha,\label{eq:unl}
\end{align}
where $ x(\tau) = 2\tau/\beta - 1 \in [-1,1]$, $y(\omega)= \omega/\wmax \in [-1,1]$.
Here, we defined
\begin{align}
u_{nl}^\alpha &\equiv \frac{1}{\sqrt{2}} \int_{-1}^{1} d x e^{\mathi \pi \{n+(1/2)\deltaF\}(x+1)} u^\alpha_l(x).\label{eq:unl}
\end{align}

\subsection{Solution of integral equation}
One can solve the integral equation in Eq.~(\ref{eq:integral-xy}) to an arbitrary precision using the procedure described in the previous study~\cite{Chikano2018}.
This is done by representing $u_l^\alpha(x)$ and $v_l^\alpha(y)$ as piece-wise polynomials and converting the original continuous problem to a discrete problem using the Galerkin method.
We refer the interested reader to Ref.~\cite{Chikano2018} for more technical details.

\section{Structure of the library}
The irbasis library consists of the following three files:
\begin{itemize}
    \item \texttt{irbasis.h5} : database file in HDF5 format
    \item \texttt{irbasis.hpp} : \verb*#C++# header file
    \item \texttt{irbasis.py} : Python script file
\end{itemize}
The first file, \texttt{irbasis.h5}, stores pre-computed basis functions, $u_l^\alpha(x)$ and $v_l^\alpha(y)$, and the singular values $s_l^\alpha$ for $\Lambda = 10, 10^2, 10^3, 10^4$.
The data structure is described later.
The rest two files, \texttt{irbasis.hpp} and \texttt{irbasis.py}, provide interfaces to the database for evaluating values of $u_l^\alpha(x)$ and $v_l^\alpha(y)$ in \verb*#C++# and Python, respectively.
The computation of other related quantities such as $u_{nl}^\alpha$ in Eq.~(\ref{eq:unl}) is also implemented.

The database has been created in the following procedure.
First, we solved the integral equation (\ref{eq:integral-xy}) in arbitrary-precision arithmetic using a Python library, irlib, developed by some of the authors~\cite{irlib_shinaoka}.
This implements the method proposed in the previous study~\cite{Chikano2018}.
Here, all the basis functions that satisfy $s_l^\alpha/s_0^\alpha > 10^{-12}$ are computed.
In practice, $u_l^\alpha(x)$ and $v_l^\alpha(y)$ are represented as piece-wise polynomials defined on multiple domains.
The degree of polynomials was chosen to be 8~\footnote{
Using a smaller degree of piece-wise polynomial substantially increases the data size and the time to solution for the integral equation because more domains are required to represent the basis functions with the desired accuracy.
}.
Solving the integral equation for the largest value of $\Lambda$ took a couple of hours on a single CPU core of a standard laptop computer.
Then, we stored the data of the piece-wise polynomial form of the basis functions and singular values in HDF5 format~\cite{hdf5_shinaoka} as double precision floating numbers.
The format of the database file is detailed in ~\ref{appendix:format}.

When using \texttt{irbasis.hpp} and \texttt{irbasis.py},
the basis functions are interpolated in double-precision arithmetic but with sufficient numerical
accuracy.
To be specific, we have confirmed that
$|\Delta u_l^\alpha(x)| / \max_{x} |u_l^\alpha(x)|$ and  $|\Delta v_l^\alpha(y)| / \max_y |v_l^\alpha(y)|$ are smaller than $10^{-8}$ at any $x$ and $y$ for all $l$.
Here, $\Delta u_l^\alpha(x)$ and $\Delta v_l^\alpha(y)$ are deviations from the reference data evaluated by arbitrary-precision mathematic.
The procedure of computing $u_{nl}^\alpha$ is detailed in~\ref{appendix:ft}.

If basis functions for any other values of $\Lambda$ are required, users can make an original database by repeating the above-mentioned procedure (see the online instruction~\cite{irbasis_github}).

\section{Installation}
The latest version of the source code, samples and the database file can be downloaded from the 
public repository \texttt{SpM-lab/irbasis} in GitHub~\cite{irbasis_github}.
The Python library \texttt{irbasis.py} depends a few standard packages: \texttt{numpy}, \texttt{h5py}, \texttt{future} (\texttt{future} is used to support both Python 2 and Python 3).
After putting \texttt{irbasis.py} and \texttt{irbasis.h5} into the working directory,
we can import irbasis from our Python project.
It is noted that irbasis libraries are included in a public repository of software for the Python programming language (PyPI)~\cite{irbasis_pypi}.
Thus, if only the python library is needed, it can be easily installed by executing the following command (the \$ sign designates a shell prompt):
\begin{verbatim}
$ pip install  irbasis 
\end{verbatim}

On the other hand, to use \texttt{irbasis.hpp},
the HDF5 C library~\cite{hdf5_shinaoka} is required. 
After installing it, we can use the \texttt{irbasis} library in our \verb*#C++# project just by including \texttt{irbasis.hpp}.
We note that the HDF5 C library must be linked to the executable at compile time.
We also provide unit tests for confirming the functionality of the \texttt{irbasis} library.
They can be run via \texttt{CTest}  and \texttt{CMake}.
More information on the usage is found on the official wiki page for \texttt{irbasis}~\cite{irbasis_github}.

\section{Basic usage}
To use the irbasis library, we first specify the dimensionless parameter $\Lambda\equiv \beta \wmax$ and the statistics. 
In practical calculations, the inverse temperature $\beta$ is usually given.
Thus, $\Lambda$ can be selected by setting the cutoff frequency $\wmax$ to a sufficiently large value such as the spectrum is bounded in $[-\wmax, \wmax]$.
After setting these parameters, the data is loaded into memory from \texttt{irbasis.h5}.
%
The following pseudo code demonstrates the usage of the interface of \texttt{irbasis.py} and \texttt{irbasis.hpp}.
\begin{lstlisting}[language=Python, basicstyle=\ttfamily\footnotesize, breaklines=true,frame=single,showstringspaces=false]
# Load basis for fermions
Lambda = 1000.0
b = load('F',  Lambda, "./irbasis.h5")

# Print all singular values
for l = 0 to b.dim()-1:
  print b.sl(l)

# Evaluate basis functions
for l = 0 to b.dim()-1:
  # Any x, y in [-1, 1]
  x = 1
  y = 1

  # u_l(x) and v_l(y)
  print b.ulx(l, x)
  print b.vly(l, y)

  # k-th derivative of u_l(x) and v_l(y)
  for k = 1, 2, 3:
    print b.d_ulx(l,x,k)
    print b.d_vly(l,y,k)

# Compute u_{ln} as a matrix for given Matsubara frequencies
nmax = 1000
unl = b.compute_unl([0, 1, ..., nmax-1])
\end{lstlisting}

You can load pre-computed data from a HDF5 file and evaluate basis functions.
We include implementations in Python (\texttt{api.py}) and \verb*#C++# (\texttt{api.cpp}) in the software package.

Figure \ref{fig:uv} shows the IR basis functions $u^\alpha_l(x)$ and $v^\alpha_l(y)$ for $\Lambda = 1000$ and $\alpha$ = F.
This figure was plotted by running the Python script \texttt{uv.py} included in the software package.
One can see that the basis functions are even/odd functions for even/odd $l$.

\section{Step-by-step examples}
This section provides step-by-step examples of how to perform typical operations in many-body calculations using the irbasis library.
For the sake of simplicity, we show only pseudocode and plot typical data.
Please refer to the sample Python/\verb*#C++# codes included in the library.
Hereafter, we omit the subscript $\alpha$ and consider only fermions ($\alpha$=F) unless otherwise stated.

As a simple but practical example,
we consider the two models defined by the spectral functions
\begin{eqnarray}
\rho(\omega) &= 
\begin{cases}
\frac{2}{\pi}\sqrt{1-\omega^2} & \mathrm{(Metal)} \label{eq:metal}\\
\frac{1}{2}\left(\delta(\omega-1) + \delta(\omega+1)\right) & \mathrm{(Insulator)} \label{eq:insuf},
\end{cases}
\end{eqnarray}
respectively.
We take $\beta=100$.

\subsection{How to compute $G_l$}
In this subsection, we explain how to compute the expansion coefficients of the Green's function $G_l$.
Since the actual procedure depends on what kind of data is available,
we discuss typical different cases below.

\subsubsection{Case 1: $\rho(\omega)$ is given.}
\begin{figure}
	\centering
	\includegraphics[width=0.45\textwidth,clip]{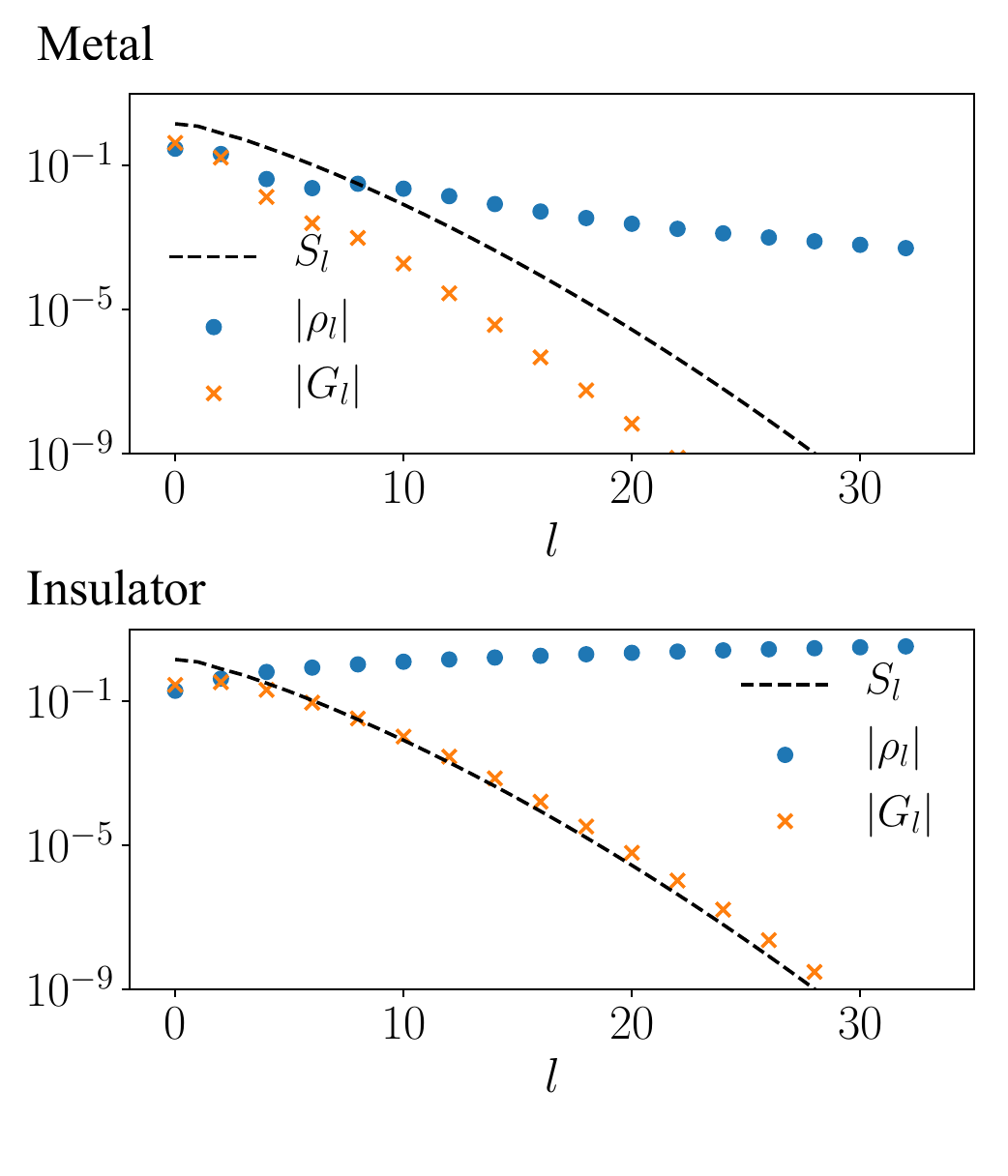}
	\caption{
		(Color online) Expansion coefficients of the spectral functions $\rho_l$ and the Green's function $G_l$ terms of IR for $\beta = 100$.
		We plot only data for even $l$.
	}
	\label{fig:glrhol}
\end{figure}

The following pseudocode demonstrates how to compute $\rho_l$ from a given spectral function $\rho(\omega)$.
What we have to do is just evaluating the integral in the right-hand side of Eq.~(\ref{eq:Gl_rhol2}).
If the spectrum consists of delta peaks, the integral can be performed analytically (see \ref{appendix:pole}).
For a continuous spectral function,
one can use an appropriate numerical integration method such as Gauss-Legendre quadrature.
Once $\rho_l$ are computed, one can readily evaluate $G_l$ using Eq.~(\ref{eq:Gl_rhol}).
Figure~\ref{fig:glrhol} shows the expansion coefficients $\rho_l$ and $G_l$ computed for the two models.

\begin{lstlisting}[language=Python, basicstyle=\ttfamily\footnotesize, breaklines=true,frame=single,showstringspaces=false]
# Cutoff frequency omega_max
omega_max = 1
  
# Transform v(y) to V(omega) (See Eq.(24))
def V(l, omega, omega_max): 
  return sqrt(1/omega_max)*b.vly(l,omega/omega_max)
  
# Transform s_l to S_l (See Eq.(25))
def S(l):      
  return sqrt(beta*omega_max/2)*b.sl(l)
  
# Compute rho_l and then G_l (See Eq.(10))
#  "integrate" denotes numerical
#  integration over omega.
for l from 0 to b.dim()-1:
  rhol = integrate(rho(omega) * V(l,omega))
  Gl = - S(l) * rhol
\end{lstlisting}


\subsubsection{Case 2: Computing $G_l$ from $G(\tau)$ by numerical integration} 
We now consider cases where $\rho(\omega)$ is unknown but $G(\tau)$ is given.
If $G(\tau)$ can be evaluated at arbitrary $\tau$ either numerically or analytically,
one can compute $G_l$ by evaluating the integral in Eq.~(\ref{eq:Gl}) numerically.
Any numerical integration scheme will suffice.
In practice, one can evaluate the integral very accurately using composite Gauss-Legendre quadrature with a small number of nodes even for large $\Lambda$ where $U_l^\mathrm{F}(\tau)$ oscillates rapidly.
See \ref{appendix:Gl} for more technical details of the composite Gauss-Legendre quadrature.
\begin{lstlisting}[language=Python, basicstyle=\ttfamily\footnotesize, breaklines=true,frame=single,showstringspaces=false]
# Transformation from u_l(x) to U_l(tau)
def U(l,tau):
  return sqrt(2/beta)*b.ulx(l,2*tau/beta-1)
  
# Compute expansion coefficients G_l
# from G(tau)
G_l = integrate(G(tau)*U(l,tau))
\end{lstlisting}

\begin{figure}
	\centering
	\includegraphics[width=0.45\textwidth,clip]{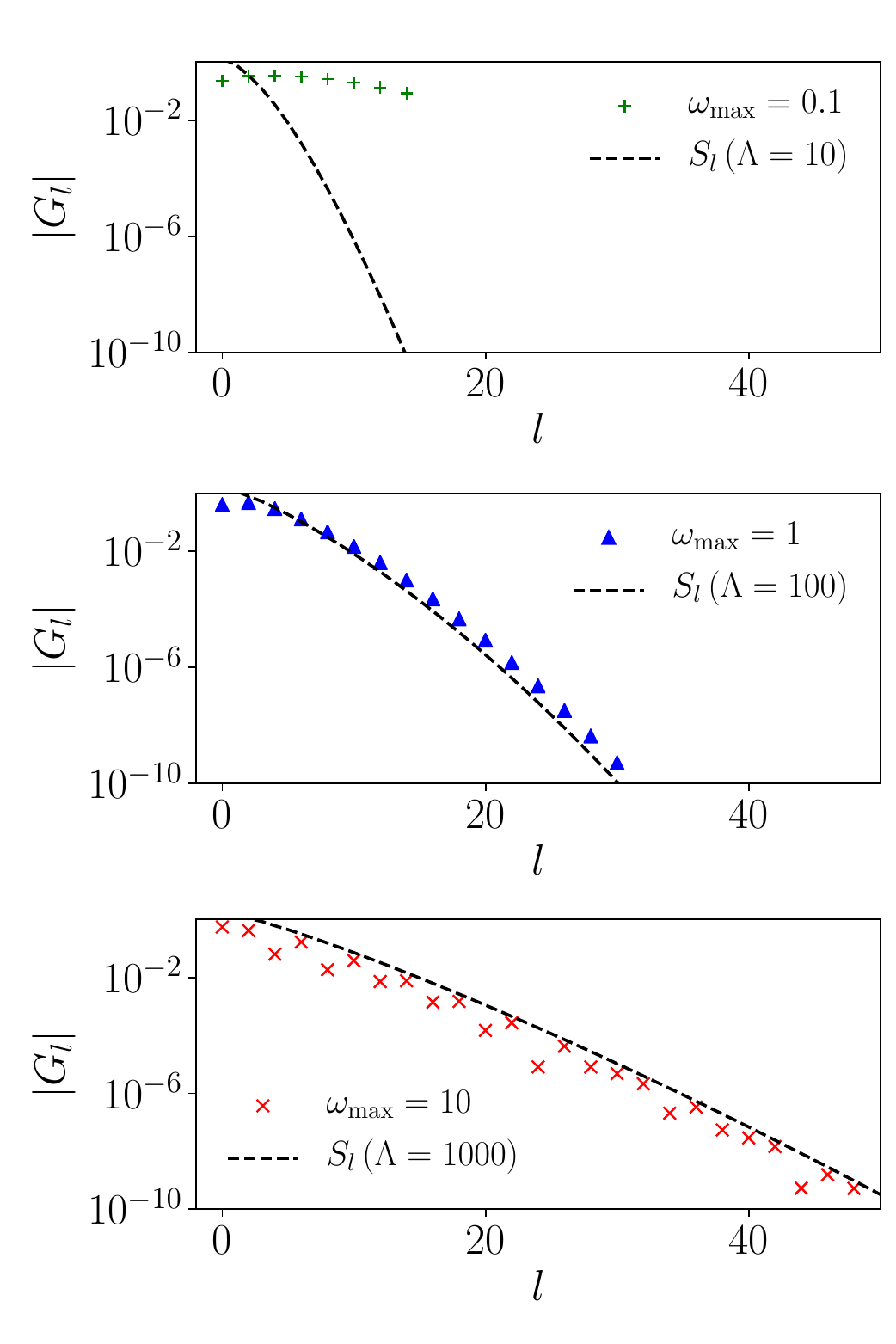}
	\caption{
		(Color online) Expansion coefficients $G_l$ computed for the insulating model with $\beta = 100$. We compare the results for $\omega_{{\rm max}} =0.1, 1, 10$.
	}
	\label{fig:opt}
\end{figure}

In general, $\wmax$ must be sufficiently large so that the spectrum function is bounded in $[-\wmax, \wmax]$.
If the exact spectrum width is unknown and only $G(\tau)$ is given,
one may have to verify whether this condition is met.
Figure~\ref{fig:opt} shows $G_l$ obtained 
by numerical integration for $\wmax = 0.1, 1, 10$ and the insulating model.
For $\wmax = 0.1$,
$G_l$ decays much more slowly than the singular values with increasing $l$, which is an indication of the violation of the condition.
When the spectrum function is well bounded (i.e., $\wmax \ge 1$),
$G_l$ decays as fast as the singular values.
As shown in the previous study~\cite{Chikano2018}, 
the number of coefficients increases only logarithmically with respect to $\wmax$.

\subsubsection{Case 3: Computing $G_l$ from $G(\tau)$ by least squares fitting}
We now consider the cases where numerical values of $G(\tau)$ are given on a predefined fine grid of $\tau$.
In such cases, one can compute $G_l$ by using linear least squares fitting techniques.

We assume that the values of $G(\tau)$ are given on a discrete grid $\{\tau_n\}$ ($n=1,2,\cdots,N_\mathrm{fit}$).
Equation~(\ref{eq:IR-decomp}) can be written in the matrix form
\begin{align}
\bm{G} = \bm{U}\bm{g},
\end{align} 
where $\bm{G}=(G(\tau_1),G(\tau_2),\cdots,G(\tau_n))^\mathrm{T},~\bm{g}= (G_1,G_2,\cdots,G_n)^\mathrm{T}, U_{lm}=U^\mathrm{F}_l(\tau_m)$.
Our task is now to compute the solution of this equation.
This can be done by minimizing
\begin{equation}
E(\bm{g}) = ||\bm{G} - \bm{U}\bm{g}||^2
\end{equation}
with respect to $\bm{g}$.
Here, $||\cdots ||$ denotes the Frobenius norm.
This procedure is stable if $N_\mathrm{fit}$ is sufficiently large so that the coefficient matrix $\bm{U}$ is well-conditioned.

\begin{figure}
	\centering
	\includegraphics[width=0.45\textwidth,clip]{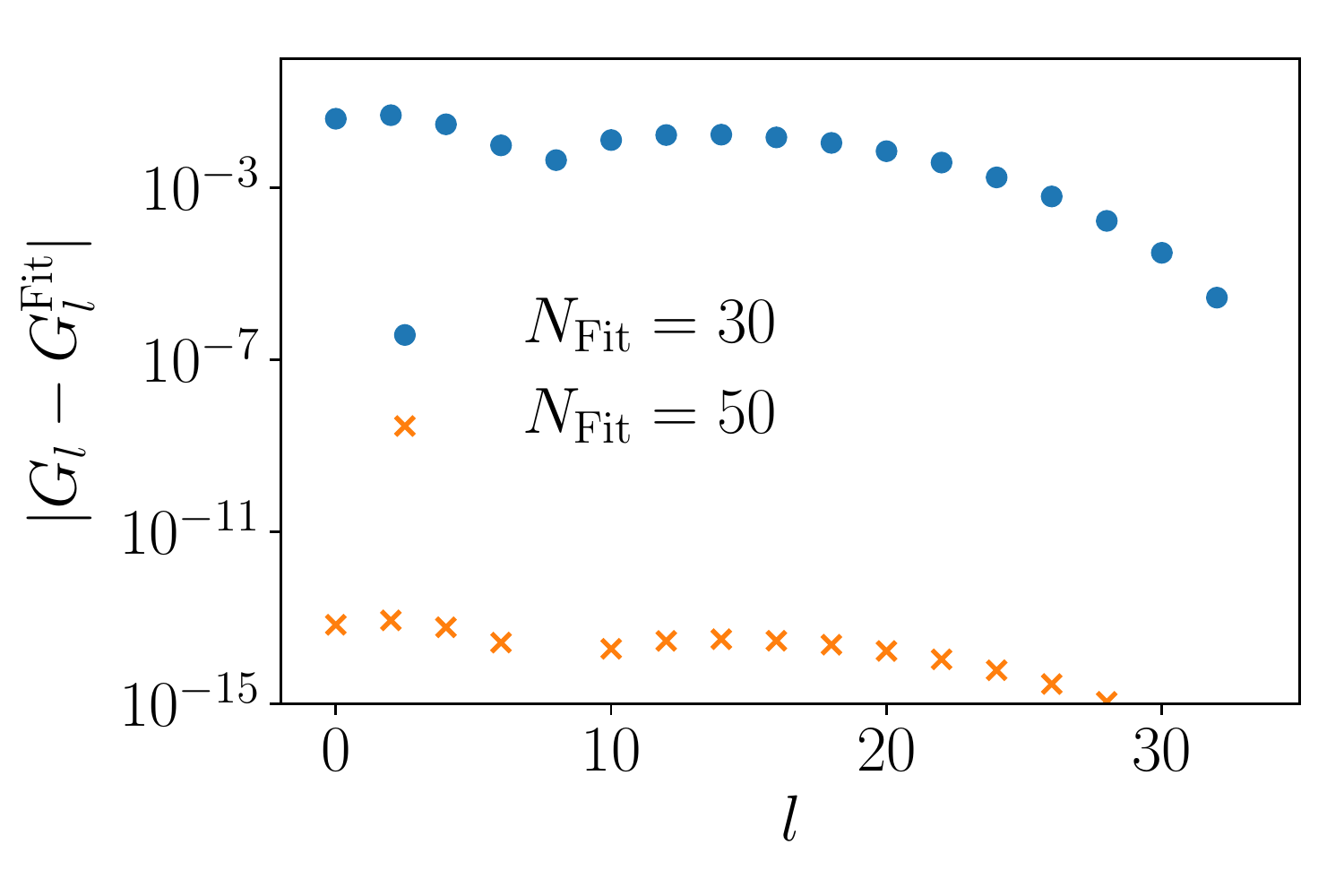}
	\caption{
		(Color online) Differences between expansion coefficients $G_l$ computed by least squares fitting and exact values for $\beta = 100$ and the insulating model.
	}
	\label{fig:fit}
\end{figure}

Let us demonstrate how the accuracy of the solution is improved as $N_\mathrm{fit}$ is increased.
In particular, we consider a uniform grid from $\tau=0$ to $\beta$.
Figure~\ref{fig:fit} shows the results computed for the insulating model.
It is clearly seen that the fit reproduces the exact results within numerical accuracy for $N_\mathrm{Fit}=50$.
Note that larger $N_\mathrm{Fit}$ is required to reconstruct accurate $G_l$ as $\beta$ increases (not shown).

\subsection{Reconstructing $G(\tau)$ and $G(i\omega_n)$ from $G_l$}
Once $G_l$ are computed,
one can evaluate $G(\tau)$ and $G(i\omega_n)$ at an arbitrary $\tau$ or $\omega_n$.
It should be noted that the basis functions included in this library are given in dimensionless form.
Thus, one have to use Eqs.~(\ref{eq:Ul})--(\ref{eq:unl}) to convert the basis functions in dimensionless form to $U^\mathrm{F}_l(\tau)$ and $U^\mathrm{F}(i\omega_n)$.

The following pseudocode describes how to use the irbasis to perform this transformation.
For simplicity, we take $\tau=\beta/2$, $n=0, \cdots, 4$ in the pseudocode.
\begin{lstlisting}[language=Python, basicstyle=\ttfamily\footnotesize, breaklines=true,frame=single,showstringspaces=false]
# Evaluate G(tau)
tau = beta/2
Gtau = 0
for l = 0 to b.dim()-1:
  Gtau += Gl[l] * U(l, tau) 

# Evaluate G(iw_n) for n = 0, ..., 4
#  "multiply" denotes matrix-vector multiplication.
#  Giwn is a vector.
n = [0, 1, 2, 3, 4]
unl = b.compute_unl(n)
Giwn = sqrt(beta) * multiply(unl, Gl)
\end{lstlisting}

\begin{figure}
	\centering
	\includegraphics[width=0.45\textwidth,clip]{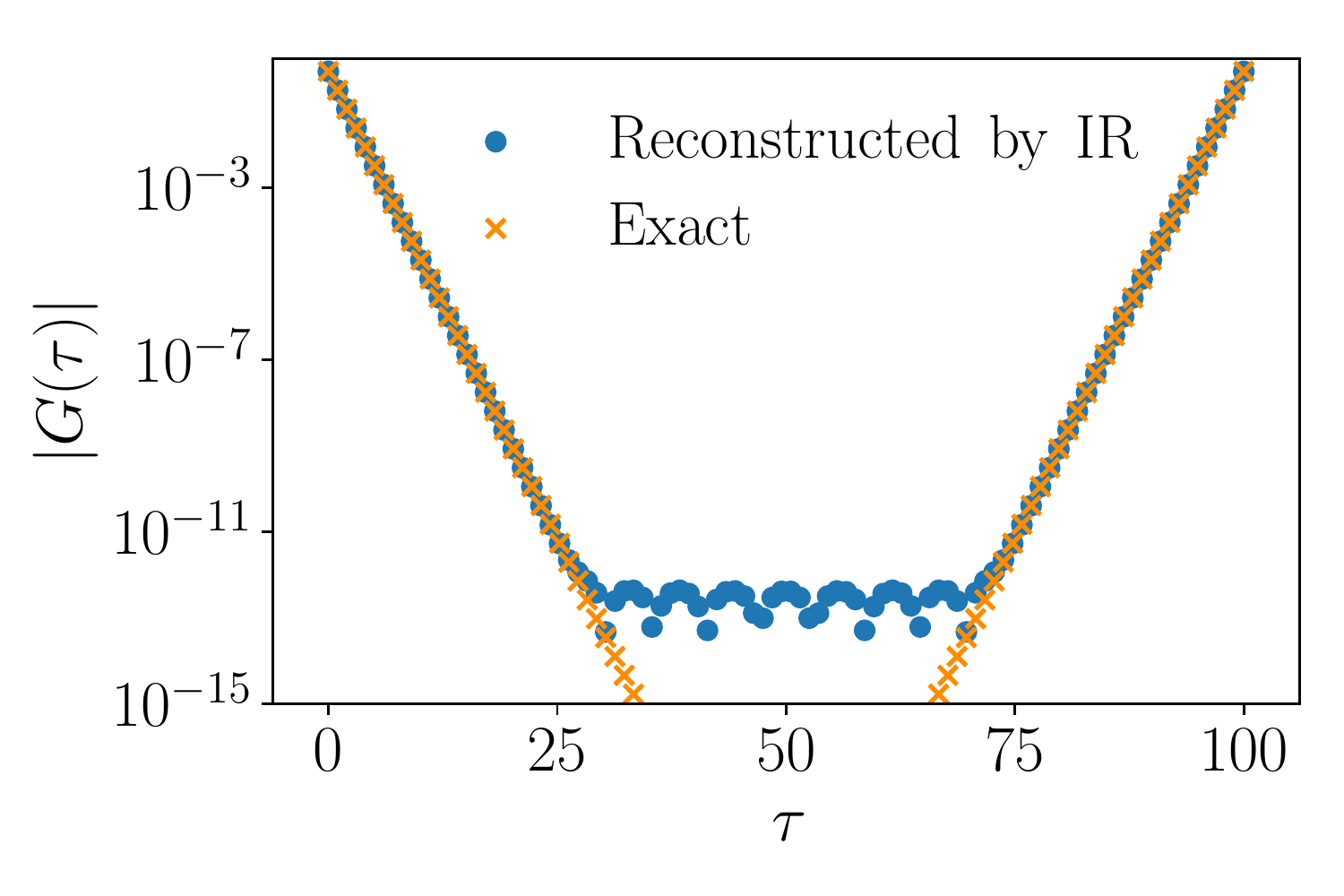}
	\includegraphics[width=0.45\textwidth,clip]{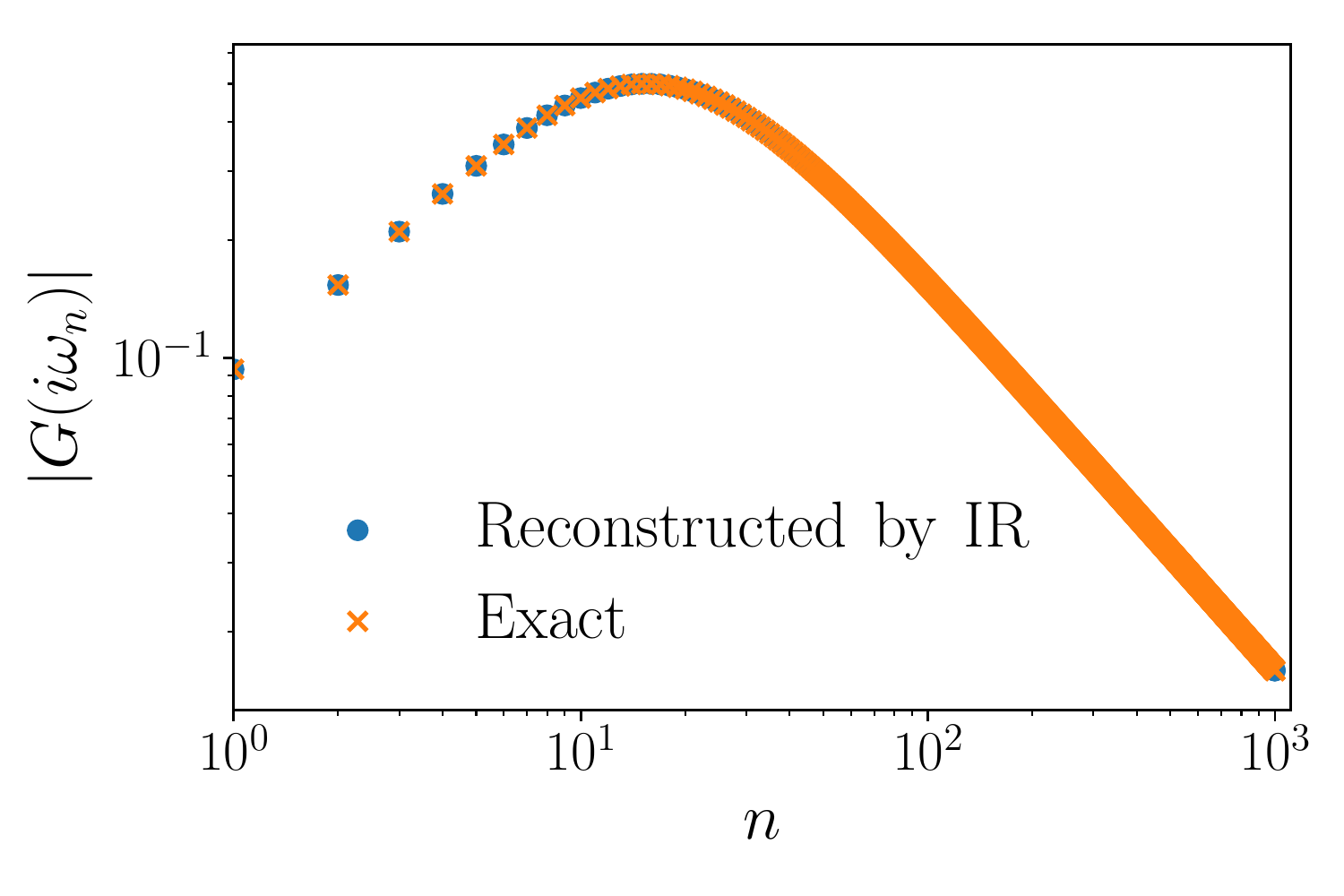}
	\caption{
		(Color online) $G(\tau)$ and $G(i\omega_n)$ reconstructed from $G_l$ for the insulating model and $\beta = 100$.
	}
	\label{fig:gtau-reconst}
\end{figure}

Figure~(\ref{fig:gtau-reconst}) shows $G(\tau)$ and $G(i\omega_n)$ reconstructed from $G_l$ for the insulating model.
We compare the reconstructed values with exact ones.
Note that the analytic form of $G(i\omega_n)$ is given by
\begin{equation}
G(i\omega_n) = \frac{1}{2}\left(\dfrac{1}{i\omega_n +1}+\dfrac{1}{i\omega_n -1}\right),
\end{equation}
where $\omega_n = (2n+1)\pi/\beta$.
One can see perfect agreement for both of $G(i\omega_n)$  and $G(\tau)$.

\section{Summary}

The IR basis yields an extremely compact representation of the Matsubara Green's function.
To obtain the IR basis, however, we need to solve an integral equation, 
which prevents the IR basis from coming into practical use in many applications.
The library, irbasis, includes pre-computed high-precision numerical solutions.
The database contains numerical solutions for typical values of the dimensional parameter $\Lambda$,
which will be sufficient for most of practical applications.
Nevertheless, users can add solutions for arbitrary values of $\Lambda$ to the database if needed.
This library allows to use the IR basis as easily as other special functions such as Legendre polynomials,
open up new and interesting research applications in computations of quantum systems.

\section*{Acknowledgments}
We thank Yuki Nagai and Markus Wallerberger for useful feedback.
HS was supported by JSPS KAKENHI Grant No. 16K17735.
HS and JO were supported by JSPS KAKENHI Grant No. 18H04301 (J-Physics).
HS, KY and JO were supported by JSPS KAKENHI Grant No. 18H01158.
KY was supported by Building of Consortia for the Development of Human Resources in Science and Technology, MEXT, Japan.

\section*{References}
\bibliographystyle{elsarticle-num}

\begin{thebibliography}{10}
\expandafter\ifx\csname url\endcsname\relax
  \def\url#1{\texttt{#1}}\fi
\expandafter\ifx\csname urlprefix\endcsname\relax\def\urlprefix{URL }\fi
\expandafter\ifx\csname href\endcsname\relax
  \def\href#1#2{#2} \def\path#1{#1}\fi

\bibitem{Georges:1996un}
A.~Georges, G.~Kotliar, W.~Krauth, M.~J. Rozenberg,
  \href{http://adsabs.harvard.edu/cgi-bin/nph-data_query?bibcode=1996RvMP...68...13G&link_type=EJOURNAL}{{Dynamical
  mean-field theory of strongly correlated fermion systems and the limit of
  infinite dimensions}}, Reviews of Modern Physics 68~(1) (1996) 13--125.
\newblock \href {http://dx.doi.org/10.1103/RevModPhys.68.13}
  {\path{doi:10.1103/RevModPhys.68.13}}.
\newline\urlprefix\url{http://adsabs.harvard.edu/cgi-bin/nph-data_query?bibcode=1996RvMP...68...13G&link_type=EJOURNAL}

\bibitem{0295-5075-100-6-67001}
J.~M. Tomczak, M.~Casula, T.~Miyake, F.~Aryasetiawan, S.~Biermann,
  \href{http://stacks.iop.org/0295-5075/100/i=6/a=67001}{Combined gw and
  dynamical mean-field theory: Dynamical screening effects in transition metal
  oxides}, EPL (Europhysics Letters) 100~(6) (2012) 67001.
\newline\urlprefix\url{http://stacks.iop.org/0295-5075/100/i=6/a=67001}

\bibitem{Hedin:1965tu}
L.~Hedin, {New method for calculating the one-particle Green's function with
  application to the electron-gas problem}, Physical Review 139~(3A) (1965)
  A796.

\bibitem{doi:10.1021/ct300648t}
M.~J. van Setten, F.~Weigend, F.~Evers,
  \href{https://doi.org/10.1021/ct300648t}{The gw-method for quantum chemistry
  applications: Theory and implementation}, Journal of Chemical Theory and
  Computation 9~(1) (2013) 232--246, pMID: 26589026.
\newblock \href {http://arxiv.org/abs/https://doi.org/10.1021/ct300648t}
  {\path{arXiv:https://doi.org/10.1021/ct300648t}}, \href
  {http://dx.doi.org/10.1021/ct300648t} {\path{doi:10.1021/ct300648t}}.
\newline\urlprefix\url{https://doi.org/10.1021/ct300648t}

\bibitem{0034-4885-61-3-002}
F.~Aryasetiawan, O.~Gunnarsson,
  \href{http://stacks.iop.org/0034-4885/61/i=3/a=002}{The gw method}, Reports
  on Progress in Physics 61~(3) (1998) 237.
\newline\urlprefix\url{http://stacks.iop.org/0034-4885/61/i=3/a=002}

\bibitem{Rubtsov:2005iwa}
A.~Rubtsov, V.~Savkin, A.~Lichtenstein,
  \href{http://link.aps.org/doi/10.1103/PhysRevB.72.035122}{{Continuous-time
  quantum Monte Carlo method for fermions}}, Physical Review B 72~(3) (2005)
  035122.
\newblock \href {http://dx.doi.org/10.1103/PhysRevB.72.035122}
  {\path{doi:10.1103/PhysRevB.72.035122}}.
\newline\urlprefix\url{http://link.aps.org/doi/10.1103/PhysRevB.72.035122}

\bibitem{Werner:2006ko}
P.~Werner, A.~Comanac, L.~de{\textquoteright} Medici, M.~Troyer, A.~Millis,
  \href{http://link.aps.org/doi/10.1103/PhysRevLett.97.076405}{{Continuous-Time
  Solver for Quantum Impurity Models}}, Physical Review Letters 97~(7) (2006)
  076405.
\newblock \href {http://dx.doi.org/10.1103/PhysRevLett.97.076405}
  {\path{doi:10.1103/PhysRevLett.97.076405}}.
\newline\urlprefix\url{http://link.aps.org/doi/10.1103/PhysRevLett.97.076405}

\bibitem{Gull:2008cma}
E.~Gull, P.~Werner, O.~Parcollet, M.~Troyer,
  \href{http://stacks.iop.org/0295-5075/82/i=5/a=57003?key=crossref.a04bd39c153e80d2afe29b4a20da2527}{{Continuous-time
  auxiliary-field Monte Carlo for quantum impurity models}}, EPL (Europhysics
  Letters) 82~(5) (2008) 57003.
\newblock \href {http://dx.doi.org/10.1209/0295-5075/82/57003}
  {\path{doi:10.1209/0295-5075/82/57003}}.
\newline\urlprefix\url{http://stacks.iop.org/0295-5075/82/i=5/a=57003?key=crossref.a04bd39c153e80d2afe29b4a20da2527}

\bibitem{Otsuki:2007ff}
J.~Otsuki, H.~Kusunose, P.~Werner, Y.~Kuramoto,
  \href{http://journals.jps.jp/doi/10.1143/JPSJ.76.114707}{{Continuous-Time
  Quantum Monte Carlo Method for the Coqblin{\textendash}Schrieffer Model}},
  Journal of the Physical Society of Japan 76~(11) (2007) 114707--11.
\newblock \href {http://dx.doi.org/10.1143/JPSJ.76.114707}
  {\path{doi:10.1143/JPSJ.76.114707}}.
\newline\urlprefix\url{http://journals.jps.jp/doi/10.1143/JPSJ.76.114707}

\bibitem{Gull:2011jda}
E.~Gull, A.~J. Millis, A.~I. Lichtenstein, A.~N. Rubtsov, M.~Troyer, P.~Werner,
  \href{http://link.aps.org/doi/10.1103/RevModPhys.83.349}{{Continuous-time
  Monte~Carlo methods for quantum impurity models}}, Reviews of Modern Physics
  83~(2) (2011) 349--404.
\newblock \href {http://dx.doi.org/10.1103/RevModPhys.83.349}
  {\path{doi:10.1103/RevModPhys.83.349}}.
\newline\urlprefix\url{http://link.aps.org/doi/10.1103/RevModPhys.83.349}

\bibitem{Boehnke2011}
L.~Boehnke, H.~Hafermann, M.~Ferrero, F.~Lechermann, O.~Parcollet, {Orthogonal
  polynomial representation of imaginary-time Green's functions}, Physical
  Review B - Condensed Matter and Materials Physics 84~(7) (2011) 1--13.
\newblock \href {http://arxiv.org/abs/1104.3215} {\path{arXiv:1104.3215}},
  \href {http://dx.doi.org/10.1103/PhysRevB.84.075145}
  {\path{doi:10.1103/PhysRevB.84.075145}}.

\bibitem{Gull2018}
E.~Gull, S.~Iskakov, I.~Krivenko, A.~A. Rusakov, D.~Zgid,
  \href{http://arxiv.org/abs/1805.03521}{{Chebyhsev polynomial representation
  of imaginary time response functions}} (2018) 1--11\href
  {http://arxiv.org/abs/1805.03521} {\path{arXiv:1805.03521}}.
\newline\urlprefix\url{http://arxiv.org/abs/1805.03521}

\bibitem{Shinaoka2017}
H.~Shinaoka, J.~Otsuki, M.~Ohzeki, K.~Yoshimi, {Compressing Green's function
  using intermediate representation between imaginary-time and real-frequency
  domains}, Physical Review B 96~(3) (2017) 1--8.
\newblock \href {http://arxiv.org/abs/1702.03054} {\path{arXiv:1702.03054}},
  \href {http://dx.doi.org/10.1103/PhysRevB.96.035147}
  {\path{doi:10.1103/PhysRevB.96.035147}}.

\bibitem{Shinaoka2018}
H.~Shinaoka, J.~Otsuki, K.~Haule, M.~Wallerberger, E.~Gull, K.~Yoshimi,
  M.~Ohzeki, {Overcomplete compact representation of two-particle Green's
  functions}, Physical Review B 97~(20) (2018) 1--14.
\newblock \href {http://arxiv.org/abs/1803.01916} {\path{arXiv:1803.01916}},
  \href {http://dx.doi.org/10.1103/PhysRevB.97.205111}
  {\path{doi:10.1103/PhysRevB.97.205111}}.

\bibitem{Chikano2018}
N.~Chikano, J.~Otsuki, H.~Shinaoka,
  \href{https://link.aps.org/doi/10.1103/PhysRevB.98.035104}{Performance
  analysis of a physically constructed orthogonal representation of
  imaginary-time green's function}, Phys. Rev. B 98 (2018) 035104.
\newblock \href {http://dx.doi.org/10.1103/PhysRevB.98.035104}
  {\path{doi:10.1103/PhysRevB.98.035104}}.
\newline\urlprefix\url{https://link.aps.org/doi/10.1103/PhysRevB.98.035104}

\bibitem{Otsuki:2017er}
J.~Otsuki, M.~Ohzeki, H.~Shinaoka, K.~Yoshimi,
  \href{http://link.aps.org/doi/10.1103/PhysRevE.95.061302}{{Sparse modeling
  approach to analytical continuation of imaginary-time quantum Monte Carlo
  data}}, Physical Review E 95~(6) (2017) 061302(R)--6.
\newblock \href {http://dx.doi.org/10.1103/PhysRevE.95.061302}
  {\path{doi:10.1103/PhysRevE.95.061302}}.
\newline\urlprefix\url{http://link.aps.org/doi/10.1103/PhysRevE.95.061302}

\bibitem{Nagai2018}
Y.~Nagai, H.~Shinaoka, {Converged spectrum in the exact-diagonalization-based
  dynamical mean-field theory : Sparse modeling analysis with the intermediate
  representation}~(5) (2018) 5--10.
\newblock \href {http://arxiv.org/abs/arXiv:1806.10316}
  {\path{arXiv:arXiv:1806.10316}}.

\bibitem{irlib_shinaoka}
\url{https://github.com/SpM-lab/irlib}.

\bibitem{hdf5_shinaoka}
\url{http://www.hdfgroup.org/HDF5}.

\bibitem{irbasis_github}
\url{https://github.com/SpM-lab/irbasis}.

\bibitem{irbasis_pypi}
\url{https://pypi.org/project/irbasis}.

\end{thebibliography}

\appendix
\section{Format of datafile}\label{appendix:format}
In a database file, we store $u^\alpha_l(x)$ and $v^\alpha_l(y)$ as piece-wise polynomial defined in the interval $[0,1]$ since these functions are even or odd.
A piece-wise polynomial $f(x)$ on $[0,1]$ is represented in the form
\begin{align}
  f(x) &= \sum_{s=0}^{N_\mathrm{s}-1} \sum_{k=0}^{N_p-1} W(x, x_s, x_{s+1}) f_{sk} (x-x_s)^k,\label{eq:pp}
\end{align}
where $f_{sk}$ is a coefficient of this piece-wise polynomial.
$N_s$ is a total number of points where $f(x\in [0, 1])=0$. 
The set $\{x_0~(=0), x_1, \cdots, x_{N_\mathrm{s}}(=1)\}$ is ascending order and $f(x)$ is equal to be $0$ at each points.
Here, $N_p$ is the degree of the piece-wise polynomial and $W(x, a, b)$ is the window function defined as
\begin{align}
W(x, a, b) &\equiv
    \begin{cases}
    1 & a \le x < b,\\
    0 & \mathrm{otherwise}.
    \end{cases}
\end{align}
Equation~(\ref{eq:pp}) is evaluated at $x=1$ as $f(1) = f(1 + 0^-)$.
We use Eq.~(\ref{eq:pp}) to evaluate the values of IR basis functions for arbitrary values of $x$ and $y$.
Derivatives of the basis functions are evaluated by differentiating Eq.~(\ref{eq:pp}).

The file format of the database used in irbasis is HDF5.
IR basis sets for different parameter sets are stored in different HDF5 groups.
Each group must have the structure given in Table~\ref{tb: db}. 
\begin{table*}[httb]
  \begin{tabular}{|l|l|l|} \hline
    Dataset & Type and dimensions & Description\\ \hline 
    info/Lambda & \textit{double} & Dimensionless parameter $\Lambda$. \\ \hline
    info/dim & \textit{int} &  Number of basis functions and singular values (\textit{dim}). \\ \hline
    info/statistics & \textit{int} & Statistics (1 for fermions and 0 for bosons).        \\ \hline
    sl & \textit{double} (\textit{dim})&  Singular values $s_l^{\alpha}$. \\  \hline 
    ulx/np& \textit{int} & Degree of the piece-wise polynomials for $u_{l}^{\alpha}(x)$ ($=N_\mathrm{p}^u$).   \\ \hline
    ulx/ns& \textit{int} & Number of sections for $x \in [0,1]$ ($=N_\mathrm{s}^u$).       \\ \hline
    ulx/data &\textit{double} (\textit{dim},~$N_\mathrm{s}^u$, $N_\mathrm{p}^u$)& Coefficients of the piece-wise polynomials for $u_l^{\alpha}(x)$.    \\ \hline
    vly/np& \textit{int} & Degree of the piece-wise polynomials for $v_{l}^{\alpha}(y)$ ($=N_\mathrm{p}^v$).   \\ \hline
    vly/ns& \textit{int} & Number of sections for $y \in [0, 1]$ ($=N_\mathrm{s}^v$).       \\ \hline
    vly/data &\textit{double} (\textit{dim},~$N_\mathrm{s}^v$, $N_\mathrm{p}^v$) & Coefficients of the piece-wise polynomials for $v_l^{\alpha}(y)$. \\ \hline
  \end{tabular}
  \caption{The structure of HDF5 group for storing data of an IR basis set.
  The actual data types are \texttt{H5T\_IEEE\_F64LE} and \texttt{H5T\_STD\_I64LE} for \textit{double} and \textit{int}, respectively.
  }
\label{tb: db}
\end{table*}

\section{Fourier transformation}\label{appendix:ft}
Equation~(\ref{eq:unl}) is evaluated by means of numerical integration over $x$ or a high-frequency expansion for low and high frequencies, respectively.

The high frequency expansion of $u_{nl}^\alpha$ is given as
\begin{align}
u_{nl}^\mathrm{F} &= \sum_{m=0}^{N_P-1} \left[\frac{-1}{i\pi (n+1/2)}\right]^{m+1} \frac{u_l^{(m), F}(1)+u_l^{(m), F}(-1)}{\sqrt 2},\\
u_{nl}^\mathrm{B} &= \sum_{m=0}^{N_P-1} \left[\frac{1}{i\pi (n+1)}\right]^{m+1} \frac{u_l^{(m), B}(1)-u_l^{(m), B}(-1)}{\sqrt 2},
\end{align}
where $u_l^{(m), \alpha}(x)$ is the $m$-th derivative function of $u_l^\alpha(x)$.
The evaluation of these equations for high frequencies is efficient and stable as long as the expansion is well converged with respect to $m$.

At low frequencies, we evaluate Eq.~(\ref{eq:unl}) by means of numerical integration for each segment (domain) of the piece-wise polynomials as
\begin{align}
u_{nl}^\alpha &= \frac{1}{\sqrt{2}} e^{\mathi \Omega_n^\alpha } \sum_{s=0}^{N_\mathrm{s}-1} J_0(x_{s}, x_{s+1}, \Omega_n^\alpha).
\end{align}
where $\Omega_n^\alpha = \pi (n + (1/2) \delta_{\alpha,\mathrm{F}})$, $N_\mathrm{s}$ is the number of segments of the piece-wise polynomials for $u_l^\alpha(x)$.
The end points of segments are denoted by $x_s$ ($x_s < x_{s+1}$).
To simplify the equation, we also define
\begin{align}
    J_k(x_s, x_{s+1}, \Omega) &\equiv \int_{x_s}^{x_{s+1}} d x e^{i \Omega x} u^{(k),\alpha}(x).\label{eq:Jk}
\end{align}

Each term is evaluated using either of the two different methods described below.
For $\Omega_n^\alpha (x_{s+1}-x_s) < c \pi$ ($c$ is a constant of $O(1)$),
Equation~(\ref{eq:Jk}) is evaluated precisely by means of numerical integration using
a high-order Gauss Legendre quadrature formula.
For $\Omega_n^\alpha (x_{s+1}-x_s) > c \pi$, we use the following recursion relation
\begin{align}
& J_k(x_s, x_{s+1}, \Omega) = \frac{1}{i \Omega} [ e^{i\Omega x_{s+1}} f^{(k)}(x_{s+1})  \nonumber\\
& - e^{i \Omega x_s} f^{(k)}(x_s) - J_{k+1}(x_s, x_{s+1}, \Omega) ]
\end{align}
with $J_{N_P}(x_s, x_{s+1}, \Omega) = 0$.

There are several useful relations between the matrix elements of $u_{nl}^\alpha$:
\begin{align}
\sum_{n\in Z} u_{nl}^* u_{nl^\prime} &= \delta_{ll^\prime},\\
(u_{nl}^\mathrm{F})^* &= u_{-n-1,l}^\mathrm{F} = (-1)^{l+1} u_{nl}^\mathrm{F},\\
(u_{nl}^\mathrm{B})^* &= u_{-n,l}^\mathrm{B} = (-1)^{l} u_{nl}^\mathrm{B}.
\end{align}

\section{Computing expansion coefficients using numerical integration}\label{appendix:Gl}
For given $G(\tau)$,
the expansion coefficients in terms of IR $G_l$ can be computed by means of numerical integration as
\begin{align}
    G_l &= \int_0^\beta d \tau G(\tau) U_l^\mathrm{F}(\tau)\\
    &= \sqrt{\frac{\beta}{2}} \int_{-1}^1 d x~G(\beta (x+1)/2) u_l^\mathrm{F}(x)\\
    &= \sqrt{\frac{\beta}{2}} \int_{-1}^1 d x~G(\beta (x+1)/2) u_l^\mathrm{F}(x).\label{eq:Gl-exp}
\end{align}
Here, the integrand functions $G(\beta (x+1)/2) u_l^\mathrm{F}(x)$ are rapidly oscillating functions around $x = \pm 1$.
There integrals can be evaluated precisely by means of composite Gauss-Legendre quadrature as detailed below.

Let us first introduce Gauss-Legendre quadrature.
An $n$-point Gauss-Legendre quadrature rule reads
\begin{align}
    \int_{-1}^1 d x f(x) \simeq \sum_{i=1}^N w^\mathrm{GL}_i f(x^\mathrm{GL}_i),\label{eq:GL}
\end{align}
where the samplings points $x^\mathrm{GL}_i$ are the roots (zeros) of the $N$-th Legendre polynomial $P_N(x)$.
The weights $w^\mathrm{GL}_i$ are $w^\mathrm{GL}_i = 2/[(1-(x^\mathrm{GL}_i)^2)(P^\prime_N(x_i^\mathrm{GL}))^2]$.
The approximate equality becomes exact if the integrand function $f(x)$ is a polynomial of degree $2N-1$ or less.
Equation~(\ref{eq:GL}) still produces a good estimate if $f(x)$ is approximated by a polynomial of degree $2N-1$.
In practice, $N$ is increased until convergence is reached.
The computation of the sampling points and weights are implemented in many numerical libraries such as \texttt{numpy}, \texttt{GSL}, \texttt{QUADPACK}.

In the present case, the basis functions $u_l^\mathrm{F}(x)$ are approximated well by piece-wise polynomials of fixed degree in each section of the piece-wise polynomials.
Thus, Equation~(\ref{eq:Gl-exp}) is approximated accurately 
\begin{align}
    G_l &= \sqrt{\frac{\beta}{2}}
    \sum_{s=0}^{N_\mathrm{s}-1}
    \sum_{i=1}^N \left(\frac{x_{s+1} - x_s}{2} w^\mathrm{GL}_i\right) \nonumber\\
    &\hspace{1em} \times  G(\beta (x_{s,i}+1)/2) u_l^\mathrm{F}(x_{s,i})\\
    &= \sqrt{\frac{\beta}{2}}\sum_{j=1}^{N_\mathrm{s}N} \tilde{w}_j
    G(\beta (x_{j}+1)/2) u_l^\mathrm{F}(x_j),
\end{align}
where $x_{s, i}\equiv (x_{s+1}-x_s) (x^\mathrm{GL}_i+1)/2 + x_s$.
For simplicity of presentation, we introduced $j\equiv(s, i)$ and defined the sampling points $x_j$ and 
the weights $\tilde{w}_j\equiv(x_{s+1}-x_s)w_i^{\mathrm GL}/2$ of the \textit{composite} Gauss-Legendre quadrature.

In practice, this integral can be computed efficiently by a matrix multiplication as
\begin{align}
    G_l &= \sqrt{\frac{\beta}{2}} (\boldsymbol{G} \boldsymbol{u})_{1,l},
\end{align}
where $\boldsymbol{G}$ and $\boldsymbol{u}$ are matrices of size $(1,n N_\mathrm{s})$ and $(n N_\mathrm{s}, N_l)$ defined as
\begin{align}
    \boldsymbol{G}_{1,i} &= G(\beta (\tilde{x}_i+1)/2),\\
    \boldsymbol{u}_{i,l} &= \tilde{w}_i u_l^\mathrm{F}(\tilde{x}_i),
\end{align}
respectively.
The computational complexity is $O(n N_\mathrm{s} N_l)$.
For $\Lambda = 10^4$ and $\alpha = \mathrm{F}$ ($N_\mathrm{s}= 134$ and $N_l=61$),
the use of $n=16$ produces converged results.

\section{IR of the Green's function with a single pole}\label{appendix:pole}
We consider the Green's function with a single pole at $\epsilon$:
\begin{align}
    G^\alpha(i\omega_n) &= \frac{1}{i\omega_n - \epsilon}.
\end{align}
From Eq.~(\ref{eq:Komega-decomp}), we obtain
\begin{align}
G^{\rm{F}}(i\omega_n) &= - \sum_{l=0}^\infty S_l^{\rm{F}} U_l^{\rm{F}}(i\omega_n) V_l^{\rm{F}}(\epsilon), \\
G^{\rm{B}}(i\omega_n) &= - \sum_{l=0}^\infty S_l^{\rm{B}} U_l^{\rm{B}}(i\omega_n) V_l^{\rm{B}}(\epsilon)\epsilon^{-1}
\end{align}
for $\alpha$ = F and B, respectively.
Comparing these two equations with Eq.~(\ref{eq:Gl-omega}),
we obtain the expansion coefficients of the Green's function in terms of IR as
\begin{align}
G^{\rm{F}}_l &= -S_l^{\rm{F}}V_l^{\rm{F}}(\epsilon),\\
G^{\rm{B}}_l &= -S_l^{\rm{B}}V_l^{\rm{B}}(\epsilon) \epsilon^{-1}.\label{eq:Glb}
\end{align}

In the case of $\epsilon = 0$ for bosons,
the inverse transformation of $G^\mathrm{B}(i\omega_n) = 1/i\omega_n$ yields $G^\mathrm{B}(\tau) = -1$.
Such constant terms in the $\tau$ domain are not represented compactly in the IR basis for bosons and should be treated separately.
Please refer to Ref.~\cite{Chikano2018} for a more detailed discussion.


\end{document}